\def\sl{\slshape}

\def\VE{\vfill\eject}
   
\input BoxedEPS
\SetTexturesEPSFSpecial
\HideDisplacementBoxes

\newcommand{\ci}{\cite}

\def\cl{\centerline}
\def\bib#1{\bibitem[#1]{#1}}

\def\0#1{{(#1)}}
\def\1#1{{\hat #1}}
\def\2#1{{\tilde #1}}
\def\3#1{{\boldsymbol#1}}
\def\4#1{{\mathbb#1}}
\def\5#1{{\cal#1}}
\def\6#1{_{\scriptscriptstyle#1}}
\def\7#1{{\bar#1}}
\def\8{\infty}
\def\9#1{^{\scriptscriptstyle#1}}
\def\/#1{{\bf#1}}
\def\;#1{{\breve#1}}

\def\i{\iota}

\def\db{{d\kern-.8ex {^-}}}

\def\={\equiv}

\def\i1#1{\int_{-\infty}^\infty d#1\, }

\def\sv#1{\vskip#1ex}

\def\frame#1#2{
    \cl{\vbox{\hrule height .3pt
    \hbox{\vrule width .3pt\kern 5pt
    \vbox{\kern 5pt
    \vbox{\hsize #1cm\noindent#2}
    \kern 5pt}
    \kern 5pt\vrule width .3pt}
    \hrule height 0pt depth.3pt}}}

\documentclass[12pt]{article}   
\usepackage{amssymb,amsmath}    
\begin{document}   
\parindent=0pt

Invited paper presented at the \sl NATO Advanced
Research Workshop on Clifford Analysis and its Applications, \rm
Prague, October 30 - November 3, 2000.  

\sv4
    
\cl{\Large\bf Communication via Holomorphic Green Functions}
\sv4

{\cl{\bf  \large Gerald Kaiser\footnote{Supported by AFOSR 
    Grant  \#F49620-01-1-0271. \hfil\today}}
	 \cl{The Virginia Center for Signals and Waves}
   \cl{kaiser@wavelets.com $\bullet$\  www.wavelets.com}}

\sv5

\cl{\large\bf Abstract}
\sv3

Let $G(x_r-x_e)$ be the causal Green
function for the wave equation in four spacetime
dimensions, representing the signal received at the
spacetime point $x_r$ due to an impulse emitted at the
spacetime point $x_e$. Such emission and reception
processes are highly idealized, since no signal can be
emitted or received at a single (mathematical) point in
space and time. We present a simple model for \sl extended
\rm emitters and receivers by continuing
$G$ analytically to a function $\tilde G(z_r- z_e)$,  where
$z_e=x_e+iy_e$ is a complex spacetime point representing
a circular \sl pulsed-beam emitting antenna dish \rm
centered at $x_e$ and emitting in the direction of $y_e$, and
$z_r=x_r-iy_r$  represents a circular \sl pulsed-beam 
receiving antenna dish \rm centered at $x_r$ and receiving from
the direction of $y_r$.  The holomorphic Green function $\tilde
G(z_r- z_e)$ represents the \sl coupling \rm between the emission
from $z_e$ and the reception at $z_r$. To preserve causality and
give nonsingular coupling, the orientation vectors $y_e$ and
$y_r$ must belong to the \sl future cone \rm $V_+$ in spacetime.
Equivalently, $z_e$ and $z_r$ belong to the \sl future and past
tubes \rm in complex  spacetime, respectively. The  space
coordinates of $y_e$ and $y_r$ give the spatial orientations and
radii of the dishes, while their time coordinates determine the
\sl duration and focus \rm of the emission and reception
processes. The \sl directivity \rm $D(y)$ of the communication
process is a convex function on $V_+$, i.e., $D(y_r+y_e)\le
D(y_r)+D(y_r)$. This shows that the efficiency of the
communication can be no better than the sum of its emission
and reception components.

\VE

\section{Introduction}  

I begin by summarizing earlier work. \sl Physical wavelets
\rm were defined in \ci{K94} as wavelet-like bases for
spaces of solutions of the homogeneous wave equation \sl
(acoustic wavelets) \rm or Maxwell's equations \sl
(electromagnetic wavelets). \rm This was motivated by the
observation that information is often communicated by
acoustic or electromagnetic waves, and this fact should be
taken into account when ``processing'' the resulting
signals. All such wavelets can be obtained from a single
``mother'' wavelet by translations, scaling, rotations and
Lorentz transformations.

The construction of physical wavelets was based on a
holomorphic extension $\tilde F(x+iy)$  of solutions $F(x)$
to complex spacetime, with the imaginary spacetime
variables $y$ interpreted as singling out approximate
directions and frequencies of propagation. Thus $\tilde
F(x+iy)$ is a description of the wave intermediate between
the spacetime domain (where exact positions and times are
known but no directional or frequency information is given)
and the Fourier domain (where exact directional and
frequency information is known but no local spacetime
information is given). This is an extension to spacetime of
continuous wavelet analysis of one-dimensional \sl time
signals, \rm whose wavelet transform is intermediate
between the time domain and the frequency domain
representations.

The physical wavelets of the homogeneous wave- and Maxwell
equations were then shown to split into a sum of \sl causal
\rm and \sl anticausal \rm wavelets. Essentially, the
causal wavelets are holomorphic extensions, in the sense
of \sl positive-frequency analytic signals, \rm of the
causal (retarded) Green function, and the anticausal ones
are similar extensions of the anticausal (advanced) Green
function for the appropriate equation. The causal wavelets
are \sl pulsed-beam solutions emitted by  disk-like
sources. \rm That is, that they represent
well-directed acoustic or electromagnetic beams that are
\sl pulsed \rm in time rather than going on forever. The
direction, pulse width, and duration of these beams are
determined by the imaginary spacetime variables $y$. Such
objects have appeared previously in the engineering
literature under the name \sl complex-source pulsed beams
\rm (see Heyman and Felsen, 1989, and the references
therein).

In this paper we further develop the above analysis by
showing that the holomorphic extension of the causal Green
function describes not only the \sl emission \rm but also
the \sl reception \rm of a pulsed beam, and so
represents a \sl communication \rm between the emitting and
receiving antenna dishes.

\section{Holomorphic Green Functions}

For simplicity, we concentrate on the wave equation in
four-dimensional spacetime ${\bf R}^4$. The causal Green
function is a fundamental solution of the wave equation,
\begin{equation}
(\partial_t^2-\Delta)G({\bf x},t)=\delta({\bf x},t),
\qquad \Delta\equiv \Delta_{\bf x}\,,
\end{equation} 
given by
\begin{equation}
G({\bf x},t)=\frac{\delta(t-|{\bf x}|)}{4\pi|{\bf x}|}\,.
\end{equation} 
Its analytic extension to complex spacetime is obtained as
follows. First we extend the delta function to the
lower-half time plane by taking its positive-frequency
(analytic signal) part. This gives the 
Cauchy kernel:
\begin{equation}
\delta(t)
=\frac1{2\pi}\int_{-\infty}^{\infty}e^{-i\omega t}d\omega
\ \to\ 
\tilde\delta(\tau)=\frac1{2\pi}\int_0^{\infty}e^{-i\omega
\tau}d\omega =\frac1{2i\pi\tau},
\end{equation} 
where
\begin{equation}
\tau=t-is \quad\hbox{with}\quad s>0
\end{equation} 
is necessary for convergence. Next, we extend the
Euclidean distance $r\equiv |{\bf x}|$ to complex
space:
\begin{equation}
r=\sqrt{{\bf x}\cdot{\bf x}}\quad \to \quad
\tilde r\equiv \sqrt{{\bf z}\cdot{\bf z}}, 
\qquad {\bf z}={\bf x}-i{\bf y}\in {\bf C}^3.
\end{equation} 
Writing 
\begin{equation}
|{\bf x}|=r\quad \hbox{and}\quad |{\bf y}|=a,
\end{equation} 
we see that
\begin{equation}
\tilde r=\sqrt{r^2-a^2-2iar\cos\theta},
\end{equation}
where $\theta$ is the angle between $\bf x$ and $\bf y$.
The complex root has branch points when $r=a$ and
$\theta=0$. For fixed $\bf y$, these form a circle of
radius $a$ in the plane orthogonal to $\bf y$. In order to
make $\tilde r$ a single-valued function, we choose the
branch defined by
\begin{equation}
\Re\,\tilde r\ge 0, \ \hbox{so that}\  {\bf y}\to{\bf 0}
\ \Rightarrow \ \tilde r\to r.
\end{equation} 
The branch cut (again, for fixed $\bf y$) is then the \sl
disk\rm
\begin{equation}
S({\bf y})=\{{\bf x}: \tilde r=0\}
=\{{\bf x}: r\le a, \  \theta=0\}.
\end{equation} 
As with ordinary branch cuts in the complex plane, the disk
$S({\bf y})$ can be \sl deformed \rm continuously to a \sl
membrane, \rm as long as its boundary ($r=a,\  \theta =0$)
remains invariant. The  extended Coulomb potential 
\begin{equation}
\tilde\phi({\bf z})\equiv-\frac1{4\pi\tilde r({\bf z})}
\end{equation} 
is a  holomorphic extension of the fundamental solution
for the Laplacian  in ${\bf R}^3$. The distribution defined
by
\begin{equation}
\tilde\delta({\bf z})\equiv \Delta \tilde\phi({\bf z}),
\end{equation} 
where $\Delta$ is the distributional Laplacian with respect
to $\bf x$, is an \sl extended source distribution \rm
which contracts to the delta function as $\bf y\to 0$
\ci{K00}:
\begin{equation}
{\bf y}\to {\bf 0} \ \Rightarrow \ \tilde\delta({\bf
x}-i{\bf y}) \to \delta({\bf x}).
\end{equation} 
Since the Coulomb potential $\phi(\bf x)$ is harmonic
outside the origin, it follows that $\tilde\phi(\bf z)$ is
harmonic outside the branch disk $S(\bf y)$, and so the
distribution $\tilde\delta$ is supported on $S(\bf y)$.
Thus $S(\bf y)$ acts as an \sl extended source \rm
generalizing the usual point source of the Coulomb
potential, and this source has been constsructed simply by
analytic continuation.

We now have all the ingredients for extending the causal
Green function $G({\bf x},t)$ in (2) to complex spacetime.
To simplify the notation, denote real spacetime
points by
\begin{equation}
x=({\bf x}, t)\in {\bf R}^4, \quad
y=({\bf y}, s)\in {\bf R}^4
\end{equation}
and complex spacetime points by
\begin{equation}
z=({\bf z}, \tau)\in {\bf C}^4, \quad
{\bf z}={\bf x}-i{\bf y}\in {\bf C}^3, \quad
\tau=t-is, \ s>0,
\end{equation}
so that
\begin{equation}
z=x-iy.
\end{equation}
The \sl holomorphic Green function \rm for the wave
equation is now defined by
\begin{equation}
\tilde G(z)=\tilde G({\bf z},\tau)
=\frac{\tilde\delta(\tau-\tilde r({\bf z}))}
{4\pi\tilde r({\bf z})}
=\frac1{8i\pi^2\tilde r(\tau-\tilde r)}.
\end{equation}
Bur recall from (4) that the imaginary part of the argument
of the numerator had to be negative. 
We must therefore require that
\begin{equation}
-\Im(\tau-\tilde r)=s+\Im\,\tilde r>0.
\end{equation}
It can be shown \ci{K00} that this is equivalent to
requiring the imaginary spacetime coordinates $y=({\bf
y}, s)$ to satisfy
\begin{equation}
|{\bf y}| < s, \quad \hbox{ or }
\quad y\in
V_+,
\end{equation}
where $V_+$ is the \sl future cone \rm in spacetime. This
means that the argument $z$ of $\tilde G$ belongs to the
\sl past tube \rm ${\cal T}_-$ in complex spacetime
\ci{K94}.

\section{Pulsed-Beam Wavelets}

We now show that $\tilde G(x-iy)$ describes the emission of
a pulsed beam by an elementary ``antenna dish'' that can be
identified with the imaginary spacetime variable $y\in
V_+$, as observed at the spacetime point $x$. Note that
when $y\to 0$, this reduces to the usual interpretation of
$G(x)$ as the signal observed at $x$ due to an idealized \sl
impulse
\rm emitted at the origin.

To simplify the analysis, we suppose that the observer is
far from the source disk $S(\bf y)$ of  (9). By (7),
\begin{equation}
r\gg a\ \Rightarrow\ \tilde r\approx r-ia\cos\theta,
\end{equation}
where the choice of branch $\Re\,\tilde r \ge 0$ has been
enforced. Substituting this into (16) gives the \sl far-zone
approximation \rm
\begin{equation}
\tilde G({\bf x}-i{\bf y},t-is)
\approx\frac1{8i\pi^2 r}\cdot\frac1{t-r-iT(\theta)}\,,
\end{equation}
where
\begin{equation}
T(\theta)= s-a\cos\theta>0\quad\hbox{since}\quad ({\bf
y}, s)\in V_+\,.
\end{equation}
At a fixed position $\bf x$, (20) is easily
seen to be a \sl pulse \rm passing the observer at time
$t=r$, in accordance with causality and \sl Huygens'
principle.  \rm The \sl duration \rm of this pulse is given
by $T(\theta)$. The pulse is \sl shortest and strongest \rm
when the observer is on the front axis of the disk $S(\bf
y)$ ($\bf x$  parallel to $\bf y$), and \sl longest and
weakest \rm on the rear axis ($\bf x$  antiparallel to $\bf
y$). By making  $s-a$ small, we obtain a \sl well-focused
pulsed beam \rm concentrated around the front axis. The
smaller $s-a$, the better the focus.

Thus $y=({\bf y}, s)\in V_+$ controls the shape
of the pulsed beam $\tilde G(x-iy)$ observed at $x$. Namely,
$\bf y$ determines the radius $a=|\bf y|$ and orientation
${\bf y}/a$ of the source disk $S(\bf y)$, while $s-a$
controls the focus of the emitted pulsed beam and its
duration along the beam axis. We will label these emission
parameters by a subscript $e$:
\begin{equation}
y\to y_e\equiv({\bf y}_e, s_e)\in V_+\,.
\end{equation}
The above pulsed beam is emitted near the origin
${\bf x}=\bf0$ around the time $t=0$. To emit a pulsed beam
from any point ${\bf x}_e$ at any time $t_e$, we
need only perform a spacetime translation:
\begin{equation}
\tilde G(x-y_e)\to \tilde G(x-x_e-iy_e)=\tilde G(x-z_e),
\end{equation}
where
\begin{equation}
z_e=({\bf x}_e, t_e)+i({\bf y}_e, s_e)=x_e+iy_e
\end{equation}
belongs to the \sl future tube \rm ${\cal T}_+$ in complex
spacetime since $y_e\in V_+$.

\section{Reception and Communication of Pulsed Beams}

The holomorphic Green function $\tilde G(x-z_e)$, defined
in the past tube ${\cal T}_-$, represents a wave emitted by
an extended source described by $z_e=x_e+iy_e$, with
$x_e$ giving the spacetime coodinates of the \sl center \rm
of the source and $y_e$ giving the  \sl spacetime extension
\rm about this center (the radius and orientation of the
emitting disk, as well as the duration of the emitted
pulse). By contrast, the original Green function $G(x-x_e)$
describes an idealized \sl spherical impulse \rm emitted
from the single spacetime point $x_e$. By making the
 coordinates $x_e$ complex, we have thus obtained a
more realistic and physically interesting model for
emission.

However, our model for \sl reception \rm is still highly
idealized since the observer is supposed to measure the
pulsed beam at the single spacetime point $x$. We now
remedy this by making the observation point complex as well:
\begin{equation}
x\to z_r\equiv({\bf x}_r, t_r)-i({\bf y}_r, s_r)=x_r-iy_r\,,
\end{equation}
where we have labeled the complex \sl reception point \rm
$z_r$ with a subscript $r$. The change in sign as compared
with (24) will be explaind below.

With the formal substitution (25), we have
\begin{equation}
\tilde G(x-z_e)\to \tilde G(z_r-z_e)
=\tilde G(x_r-x_e-i(y_r+y_e)).
\end{equation}
Since the argument of $\tilde G$ must belong to the past
tube $\cal T_-$\,, we have to require that
\begin{equation}
y_r+y_e\in V_+\quad\hbox{for all}\quad y_e\in V_+\,.
\end{equation}
This implies that $y_r\in V_+$\,, which explains our choice
of sign in (25). 
\vskip 1ex

\parbox{4 in}{
\sl The emission point $z_e$ must belong to the future tube
$\cal T_+$\,, and the reception point $z_r$ must
belong to the past tube $\cal T_-$\,. 
$\tilde G(z_r-z_e)$ represents the coupling
between $z_e$ and $z_r$,  giving the strenght
of the overall communication process. \rm
}\vskip 1ex

\noindent These requirements also make intuitive sense,
since emission creates a signal in the future while
reception measures a signal from the past.
From now on we identify $z_e\in\cal T_+$ with the \sl
emitting dish \rm and $z_r\in \cal T_-$ with the \sl
receiving dish. \rm Note that this includes the \sl
durations \rm of the emission and the reception processes.
(In reception, ``duration'' is interpreted
as the \sl integration time.\rm)
Our use of the term ``dish'' therefore stretches the usual
meaning, being a \sl spacetime \rm concept rather than
merely spatial. 

The condition $z_r\in\cal T_-$ was derived from the \sl
mathematical \rm requirement that $z_r-z_e\in\cal T_-$ for
all $z_e\in \cal T_+$.  We now confirm that our model
also makes \sl physical \rm sense by studying the
communication $\tilde G(z_r-z_e)$ in the far-zone
approximation. Writing
\begin{equation}
r=|{\bf x}_r-{\bf x}_e|, 
\quad t=t_r-t_e, \quad 
a=|{\bf y}_r+{\bf y}_e|, \quad s=s_e+s_e\,,
\end{equation}
(20) gives
\begin{equation}
r\gg a\ \Rightarrow \ \tilde G(z_r-z_e)
\approx\frac1{8i\pi^2 r}\cdot\frac1{t-r-iT(\theta)},
\end{equation}
where $\theta$ is the angle between ${\bf x}_r-{\bf x}_e$
and ${\bf y}_r+{\bf y}_e$ and
\begin{equation}
T(\theta)=s-a\cos\theta
\end{equation}
now denotes the \sl duration \rm of the overall
communication process. Let us fix the distance $r$ between
the centers of the emitting and receiving disks $S({\bf
y}_e)$ and $S({\bf y}_r)$, as well as their radii $a_e=|{\bf
y}_e|$ and  $a_r=|{\bf y}_r|$ and the duration parameters
$s_e$ and $s_r$. 

To maximize the communication (29), we need to  minimize the
duration function $T(\theta)$. By Schwarz's inequality,
\begin{equation}
a\le a_r+a_e\,, \quad\hbox{with}\quad a=a_r+a_e\ \hbox{ iff
${\bf y}_r$ is \sl parallel \rm to ${\bf y}_e$\,.}
\end{equation}
Thus (30) shows that the communication is maximal when 

\begin{enumerate}

\item the emitting and receiving dishes  are synchronized
for causal communication, so that $t=r$\,;

\item the spatial direction vectors ${\bf y}_r$ and ${\bf
y}_e$ are parallel to one another (to maximize $a$) and also
parallel to 
${\bf x}_r-{\bf x}_e$ (to make $\theta=0$).

\end{enumerate}

We have seen that ${\bf y}_e$ gives the direction of
propagation of the emitted pulsed-beam wavelet. If we
similarly assume that ${\bf y}_r$ gives the direction in
which the receiving disk is pointed, the above result
clearly runs against common sense since it states that
reception is greatest when the receiver points directly \sl
away \rm from the transmitter. Rather, we must interpret
${\bf y}_r$ as a vector pointing \sl into \rm the receiver,
so that the dish $z_r=x_r-iy_r$ is configured to receive
receive signals coming \sl from \rm the direction of 
${\bf y}_r$. With this interpretation, the above results
are in complete harmony with intuition.

\vskip 1ex

\parbox{4 in}{\sl The communication  between an emitting
dish $z_e=x_e-iy_e$ and a receiving dish $z_r=x_r-iy_r$
is greatest when the two dishes are synchronized for causal
commnication and each is pointed towards the center of the
other.}

\vskip 1ex

\section{The Convex Directivity Function}

According to the above, the \sl peak value \rm of the pulsed
beam emitted by $z_e$ and received by $z_r$ is obtained
when 
${\bf y}_r, {\bf y}_e$ and ${\bf x}_r-{\bf x}_e,$ are all
parallel, so that
\begin{equation}
r=t,\ \  a=a_e+a_r,\ \ \theta =0
\end{equation}
and
\begin{equation}
\tilde G(z_r-z_e)
\approx\frac 1{8\pi^2 r}\cdot\frac1{s-a}\,.
\end{equation}
A dimensionless measure of the \sl directivity \rm of
the communication, independent of $r$, may be given
as
\begin{equation}
D\equiv \frac a{s-a}=\frac{a_r+a_e}{s_r+s_e-a_r-a_e}\,.
\end{equation}
Since this expression depends only on $y_r+y_e\in V_+$, it
defines a function $D(y)$ on $V_+$\,. Note that
\begin{equation}
0\le D(y)<\infty,\ \  \hbox{with $D(y)=0$\  iff \ $a=0$.}
\end{equation}
But under the above assumptions, $a=0$ implies ${\bf
y}_r={\bf y}_e={\bf 0}$. The directivity $D(y_r+y_e)$
therefore vanishes if and only if the emitting and 
receiving disks both shrink to points, making the
communication process entirely direction-free. (But note
that we still have $s_e>0$ and $s_r>0$, so that the
communicated signal remains a \sl pulse \rm rather than an
\sl impulse.\rm)  This helps justify the term
``directivity.''

But $D$ has another attractive property that goes deeper
than the above. For all $y_r, y_e\in V_+$ we have
$s_r-a_r>0$ and $s_e-a_e>0$, hence
\begin{equation}
D(y_r+y_e)=\frac{a_r+a_e}{(s_r-a_r)+(s_e-a_e)}\le
\frac{a_r}{s_r-a_r}+\frac{a_e}{s_e-a_e}\,,
\end{equation}
thus
\begin{equation}
D(y_r+y_e)\le D(y_r)+D(y_s).
\end{equation}
Now $V_+$ is a \sl convex cone \rm in ${\bf R}^4$, and (37)
shows that $D$ is a \sl convex function \rm on $V_+$\,.
This is an important property with an immediate physical
interpretation. $D(y_e)$ measures the directivity of
the communication when the receiver is a \sl spacetime point
\rm ($y_r=0$), so that $\tilde G(x_r-z_e)$ represents a
pure emission. Similarly, $D(y_r)$ measures the directivity
of the communication when the emitter is a spacetime point,
so that $\tilde G(z_r-x_e)$ represents a pure reception.
Then (37) states that \sl the efficiency of the overall
communication can be no better than the sum of its separate
emission and reception components. \rm Further developments
of these ideas will appear in \ci{K01}.

\end{document}